\UseRawInputEncoding
\documentclass[superscriptaddress,twocolumn,showpacs,a4paper]{revtex4-1}
\usepackage{graphicx}
\usepackage{amssymb}
\usepackage{xcolor}
\pdfoutput=1
\usepackage{ulem}
\usepackage{dcolumn}
\usepackage{bm}
\usepackage{braket}
\begin{document}

\title{Entanglement properties of superconducting qubits coupled to a semi-infinite transmission line}
	
\author{Yangqing Guo}
\author{Pingxing Chen}
	\affiliation{
		Department of Physics, National University of Defense Technology, Changsha 410073, China
	}

\author{Jian Li}
	\email{Corresponding author: lij33@sustech.edu.cn}
	\affiliation{
		Shenzhen Institute for Quantum Science and Engineering, Southern University of Science and Technology, Shenzhen 518055, China}
    \affiliation{
        International Quantum Academy, Shenzhen 518048, China}
    \affiliation{
        Guangdong Provincial Key Laboratory of Quantum Science and Engineering, Southern University of Science and Technology, Shenzhen 518055, China
	}

\date{\today}	
	
\begin{abstract}
    Quantum entanglement, a key resource in quantum information processing, is reduced by interaction between the quantum system concerned and its unavoidable noisy environment. Therefore it is of particular importance to study the dynamical properties of entanglement in open quantum systems. In this work, we mainly focus on two qubits coupled to an adjustable environment, namely a semi-infinite transmission line. The two qubits' relaxations, through individual channels or collective channel or both, can be adjusted by the qubits' transition frequencies. We examine entanglement dynamics in this model system with initial Werner state, and show that the phenomena of entanglement sudden death and revival can be observed. Due to the hardness of preparing the Werner state experimentally, we introduce a new type of entangled state called pseudo-Werner state, which preserves as much entangling property as the Werner state, and more importantly, is experiment friendly. Furthermore, we provide detailed procedures for generating pseudo-Werner state and studying entanglement dynamics with it, which can be straightforwardly implemented in a superconducting waveguide quantum electrodynamics system.
\end{abstract}	


\maketitle

\section{Introduction}
    Quantum entanglement, which expresses the quantum correlation of the subsystems in a larger quantum system, is one of the unique phenomena in quantum physics. Presently, quantum entanglement has been treated as a crucial resource required for almost all aspects of quantum information processing \cite{Nielsen2000}, {\it e.g.} quantum key distribution \cite{Ekert1991,Grosshans2003}, quantum teleportation \cite{Bennett1993,Bouwmeester1997}, dense coding \cite{Bennett1992,Mattle1996}, remote state preparation \cite{Bennett2001,Peng2003}, and quantum computation \cite{DiVincenzo2000}. However, the presence of unavoidable interaction between a quantum system and its environment, which results in relaxation and decoherence, becomes the major obstacle for manipulating entangled systems in an open quantum world. In consequence, preparation and dynamics of entangled states have been drawn much attention in the last two decades. There has been a lot of effort in studying the variation of entanglement due to decoherence in different systems, and some novel phenomena have been found, e.g. entanglement sudden death (ESD) and revival etc. \cite{Yu2004,Laurat2007,Almeida2007,Ficek2006,Li2009,Lopez2008,Das2009,Yu2006,Al2008}.

    A typical model for studying ESD \cite{Yu2004} consists of two non-interacting qubits coupling to independent environmental noises, respectively. By further replacing the two independent environmental noises with a multimode vacuum field and adding the interaction between the qubits, resurrection of the disappeared entanglement can happen \cite{Ficek2006}. The ESD phenomenon was experimentally verified with atom ensembles \cite{Laurat2007} and photon pairs \cite{Almeida2007} in 2007. After that, a system of two superconducting qubits driven by coherent microwave fields was theoretically studied, and stationary entangled states were obtained in a certain region of the parameter space \cite{Li2009}. Meanwhile, entanglement transfer from the bipartite system to the reservoirs \cite{Lopez2008}, as well as the scenario of interacted qubits contacting with different environments (reservoirs) at zero temperature \cite{Das2009} were also considered.

    Previous work has already systematically studied entanglement dynamics under different kinds of couplings between the qubits and the reservoirs theoretically, nevertheless, there is lack of setup to verify these dynamics experimentally in a single system. Very recent progress in superconducting waveguide quantum electrodynamics (QED) \cite{Kockum2018, Kannan2020, Wen2020} provides such an opportunity to experimentally investigate ESD and revival on chip. In this context, we mainly focus on exploring the entanglement dynamics with various combinations of collective and individual relaxation channels as well as qubit-qubit couplings, which can be controlled by qubits' frequencies, in a waveguide QED system. We consider a class of 'X' shape initial states (only the diagonal and anti-diagonal elements of the density matrix are non-zero), which are commonly used for studying bipartite entanglement properties.

    Our work is structured as follows. In Sec. II, we give the model and describe the kinematic equation of the system. In Sec. III, we detail the dependence of entanglement time evolution on different values of collective, individual relaxations and coupling of the qubits. In Sec. IV, we show a procedure for experimentally observing ESD and revival in a superconducting waveguide QED system. The conclusion is in Sec. V.

\section{Physical model}

\begin{figure}[bht]
    \includegraphics[width=8.5cm]{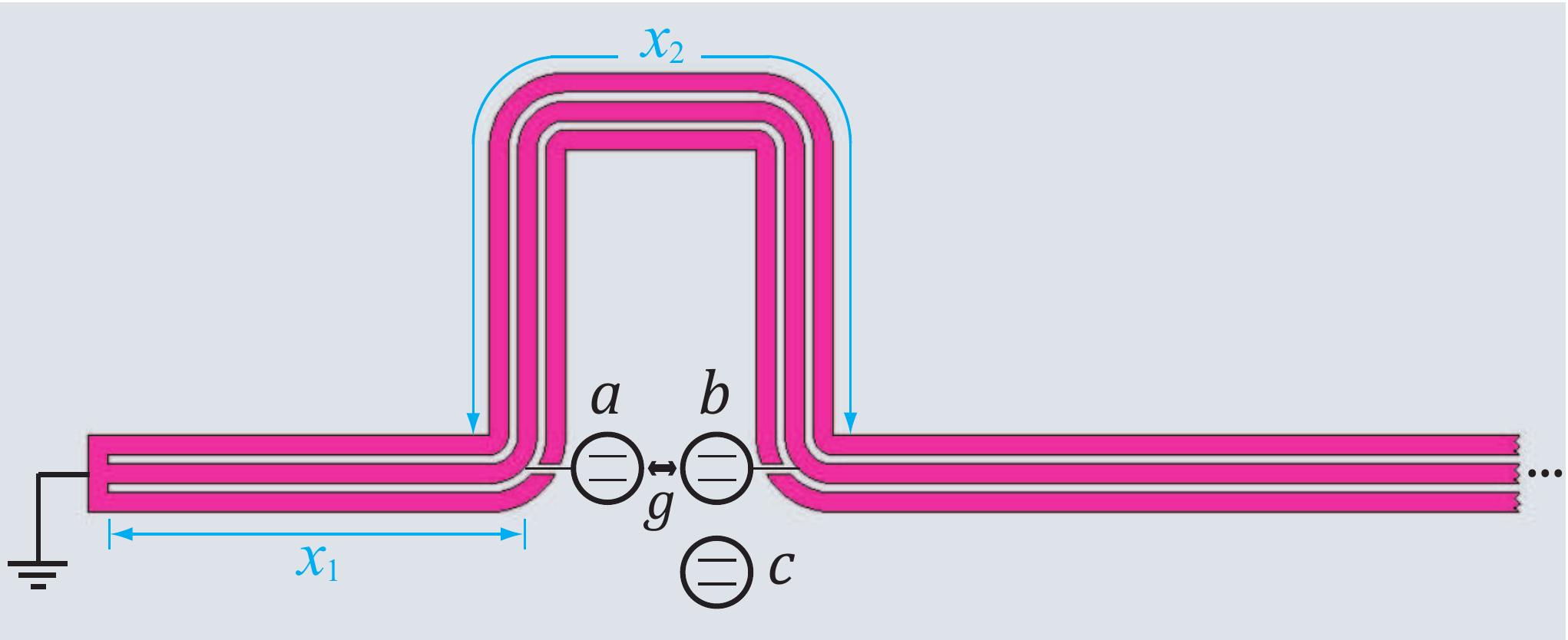}
    \caption{The schematic model of two qubits ($a$ and $b$) coupled to a semi-infinite waveguide. There is a switchable coupling $g$ between qubit-$a$ and qubit-$b$. An auxiliary qubit ($c$) is tunably coupled to qubit-$b$, and is only used for preparing initial entangled states. }
    \label{fig:1}
\end{figure}

    Here we consider a system of two small two-level atoms (qubits) coupled to a semi-infinite transmission line, very similar to the one studied in Ref.~\cite{Kockum2018}, as shown in Fig.~\ref{fig:1}. The semi-infinite transmission line has a coplanar waveguide (CPW) form, and its left-hand-side end (LHSE) is shorted to the ground. The qubits are small in comparison to the wavelengths of the CPW's bosonic modes at their transition frequencies. Thus, these two qubits ($a$ and $b$) can be considered as point-like objects. The length of the CPW between the LHSE and qubit-$a$'s coupling position is $x_1$, and the length between qubit-$a$'s and qubit-$b$'s coupling positions is $x_2$, as illustrated in Fig.~\ref{fig:1}. $x_2 = 2 x_1$ will be considered through out the following discussions. An auxiliary qubit ($c$) will be required for preparing entangled states of qubit-$a$ and qubit-$b$ in Sec. IV. However in this section, let us ignore the existence of qubit-$c$.

    If there is no direct coupling $g$ between qubit-$a$ and qubit-$b$, the system we consider here is exactly the same as the {\it small atoms $+$ mirror} model discussed in \cite{Kockum2018}. We assume that the transition frequencies of qubit-$a$ and qubit-$b$ are similar, $\omega_a \approx \omega_b$, and that the bare coupling strengths between qubit-$a,b$ and the CPW (the rates describe how fast photon can emit from qubit-$a,b$ to the CPW without any interference) are the same, $\gamma_a = \gamma_b \equiv \gamma$. We further consider that the two qubits have similar intrinsic 'non-radiative' relaxation rate $\gamma_{nr}$, which is much smaller than $\gamma$. Thus, the master equation of this system can be written as \cite{Kockum2018}

    \begin{eqnarray}
	&& \dot\rho = -i[H,\rho] + \Gamma_a {\cal D}[\sigma_-^a]\rho + \Gamma_b {\cal D}[\sigma_-^b]\rho\nonumber\\	
	&& + \Gamma_{col} \left[ \sigma_-^a\rho\sigma_+^b + \sigma_-^b\rho\sigma_+^a - \frac{1}{2} \left\{ \sigma_+^a\sigma_-^b + \sigma_+^b\sigma_-^a , \rho  \right\} \right],
	\label{master equation}
    \end{eqnarray}
    where $\Gamma_a = \gamma (1+\cos\phi) + \gamma_{nr}$ and $\Gamma_b = \gamma (1+\cos 3\phi) + \gamma_{nr}$ are individual relaxation rates of qubit-$a$ and qubit-$b$, respectively; $\Gamma_{col} = \gamma (\cos\phi + \cos 2\phi)$ is the collective relaxation rate of the two qubits. The system Hamiltonian
    \begin{eqnarray}
	H=&& \left( \omega_a + \frac{\gamma}{2}\sin\phi \right) \frac{\sigma_z^a}{2} + \left( \omega_b + \frac{\gamma}{2}\sin3\phi \right) \frac{\sigma_z^b}{2} \nonumber\\
	&&+\frac{\gamma}{2} (\sin\phi+\sin2\phi) \left(\sigma_-^a\sigma_+^b+ \sigma_+^a\sigma_-^b \right).
	\label{system_hamiltonian}
    \end{eqnarray}
    Here $\phi = 2\pi x_2 / \lambda$ denotes the phase shift between the two coupling positions of the two qubits, and $\lambda$ is the wavelength determined by qubit-$a$'s transition frequency $\omega_a$. $\sigma_{-(+)}^m$ is the lowering (raising)  operator of qubit-$m$, $\sigma_z^m$ represents the Pauli $Z$-matrix for qubit-$m$, and ${\cal D}[A]\rho=A\rho A^\dag - \frac{1}{2}\{A^\dag A,\rho\}$.

\begin{figure}[bht]
    \includegraphics[width=7cm]{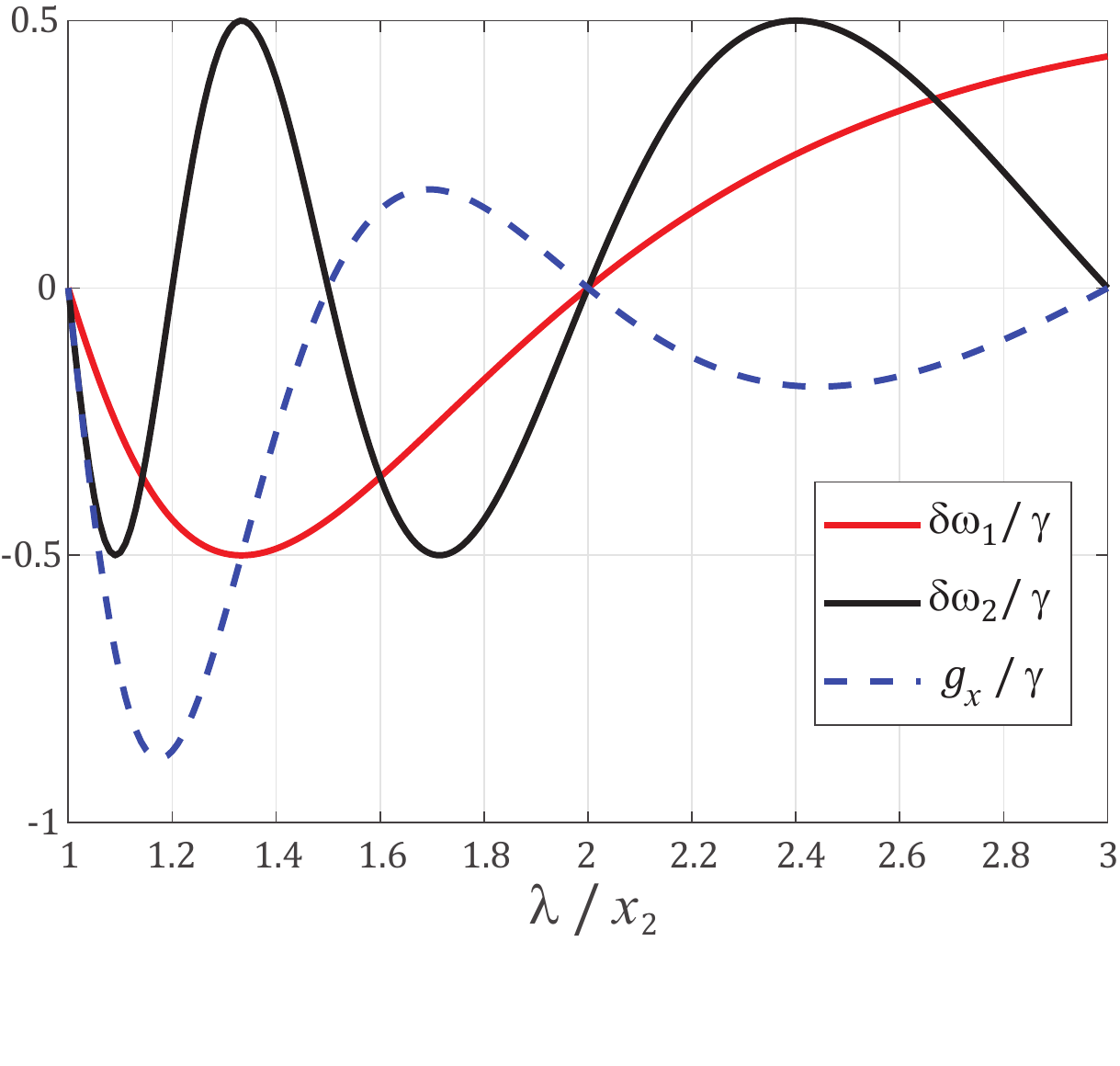}
    \caption{Qubit frequency shifts $ \delta\omega_1 $, $ \delta\omega_2 $ and exchange coupling strength $g_x$ versus wavelength $\lambda$.}
    \label{fig:2}
    \end{figure}

\begin{figure}[bht]
    \includegraphics[width=7cm]{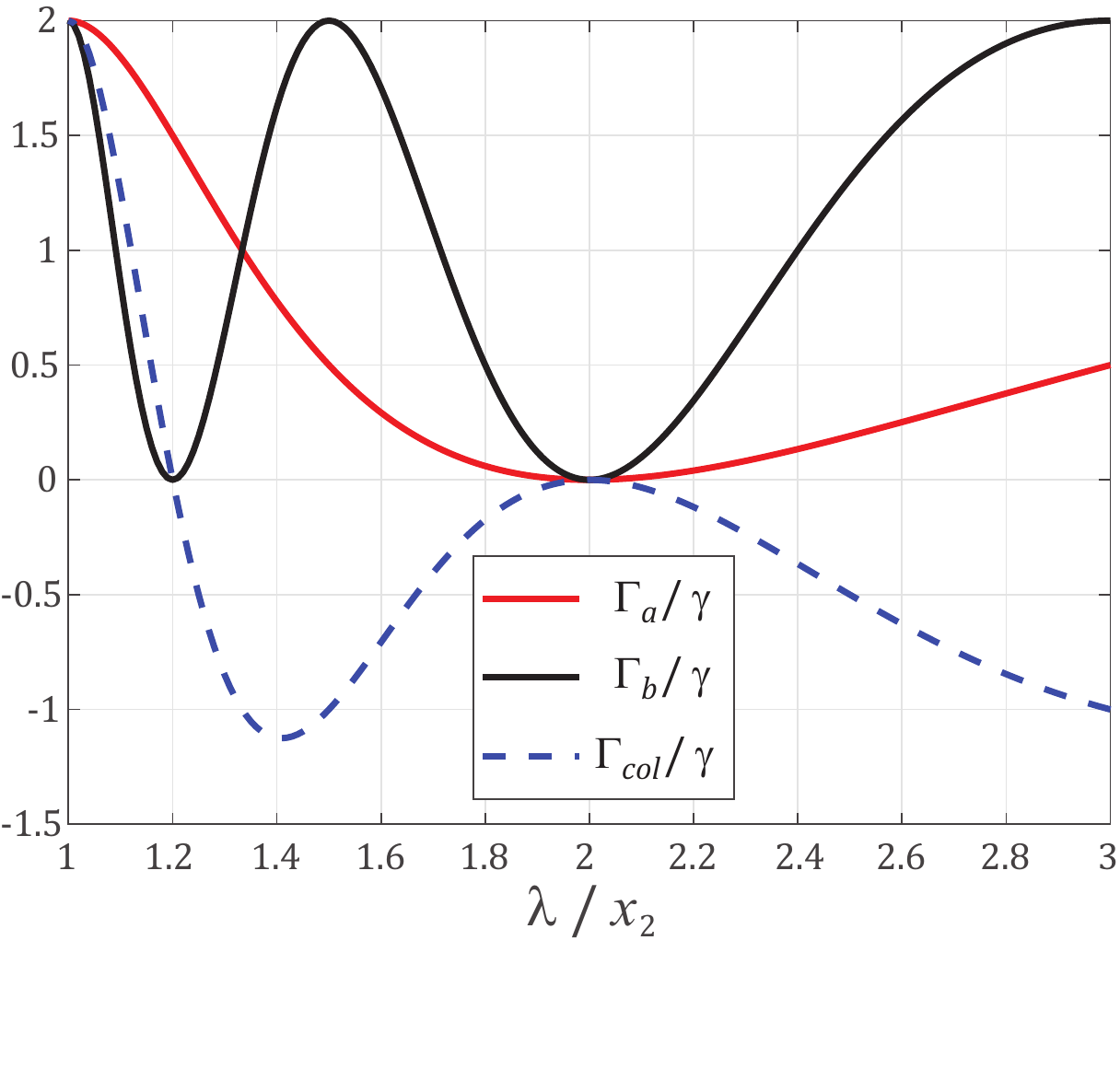}
    \caption{The individual relaxation rates $\Gamma_a$, $\Gamma_b$ and the collective relaxation rate $\Gamma_{col}$ versus wavelength $\lambda$. The intrinsic relaxation rate $\gamma_{nr} = 0$ is taken. }
    \label{fig:3}
    \end{figure}

    From the Hamiltonian Eq.~(\ref{system_hamiltonian}), one finds that couplings to the semi-infinite transmission line can induce frequency shifts $\delta\omega_1 = \frac{\gamma}{2}\sin\phi$, $\delta\omega_2 = \frac{\gamma}{2}\sin 3\phi$ for the qubits, and that a qubit-qubit exchange coupling with strength $g_x = \gamma(\sin\phi + \sin 2\phi)/2$ is also induced. Since in a waveguide QED setup, the length $x_2$ (also $x_1$) is fixed when the chip is fabricated, therefore the phase shift $\phi$ is only dependent on wavelength $\lambda$ (which is controlled by the qubit transition frequency). Thus in Fig.~\ref{fig:2}, $\delta\omega_1$, $\delta\omega_2$ and $g_x$ are plotted as functions of $\lambda$; in Fig.~\ref{fig:3}, the individual and collective relaxation rates as functions of $\lambda$ are also plotted.

    When preparing an entangled initial state of qubit-$a$ and qubit-$b$, finite coupling between the two qubits are required, and zero individual relaxation rate of each qubit is also desired for high state fidelity (the collective relaxation rate may not necessarily be zero). From Fig.~\ref{fig:3} we can find that, only at $\lambda = 2 x_2$ both individual relaxation rates $\Gamma_a$ and $\Gamma_b$ are zero (assuming the intrinsic relaxation rate $\gamma_{nr}$ is negligibly small here); but from Fig.~\ref{fig:2} we also get zero qubit-qubit exchange coupling $g_x$ at this wavelength. Thus we need an extra switchable qubit-qubit exchange coupling $g$, as shown in Fig.~\ref{fig:1}, to help generate high fidelity entangled state. In a superconducting waveguide QED system, this switchable exchange coupling can be realized by using a third qubit as a tunable coupler \cite{Yan2018}.

\section{Dynamics of entanglement}
\label{sec:3}

    For a bipartite system, Wootters' concurrence $C$ \cite{Wootters1998} is a convenient entanglement measure, especially for 'X' shape states which retain 'X' form \cite{Li2009} during the evolution governed by the master equation. Now we focus on a subclass of 'X' shape states - the Werner state \cite{Bennett1996}, which is defined as
    \begin{eqnarray}
	\rho_W &&=\frac{1-f}{3}I_4+ \frac{4f-1}{3}\ket{\Psi^-}\bra{\Psi^-} \nonumber\\
	&&=\frac{1}{3}
	\left(\begin{array}{cccc}
		1-f   & 0        & 0        & 0   \\
		0     & (1+2f)/2 & (1-4f)/2 & 0   \\
		0     & (1-4f)/2 & (1+2f)/2 & 0   \\
		0     & 0        & 0        & 1-f
	   \end{array}\right),
    \end{eqnarray}
    where $\ket{\Psi^-}=(\ket{01}-\ket{10})/\sqrt{2}$ represents the Bell singlet state, $f$ is called fidelity, and $0.25\leq f\leq1$. To solve the master equation with initial Werner state, we follow the procedure described in e.g. Ref.~\cite{Li2010} (see Appendix~\ref{x_state}).

    In the following calculations, we take the intrinsic 'non-radiative' relaxation rate $\gamma_{nr} = 2\pi\times 0.03$MHz (see Ref.~\cite{Kannan2020}), and the bare coupling strength between the qubits and the CPW $\gamma = 2\pi\times 5$MHz.

\subsection{No exchange coupling, no collective relaxation}

    By observing Fig.~\ref{fig:2} and Fig.~\ref{fig:3}, one may easily find that the most trivial wavelength is $\lambda / x_2 = 2$, at which the qubit-qubit exchange coupling $g_x = 0$, and the collective relaxation rate $\Gamma_{col} = 0$. In this case, after the initial entangled state was prepared, there is no way to induce extra entanglement, therefore only ESD can happen.

    \begin{figure}[bht]
    \includegraphics[width=7cm]{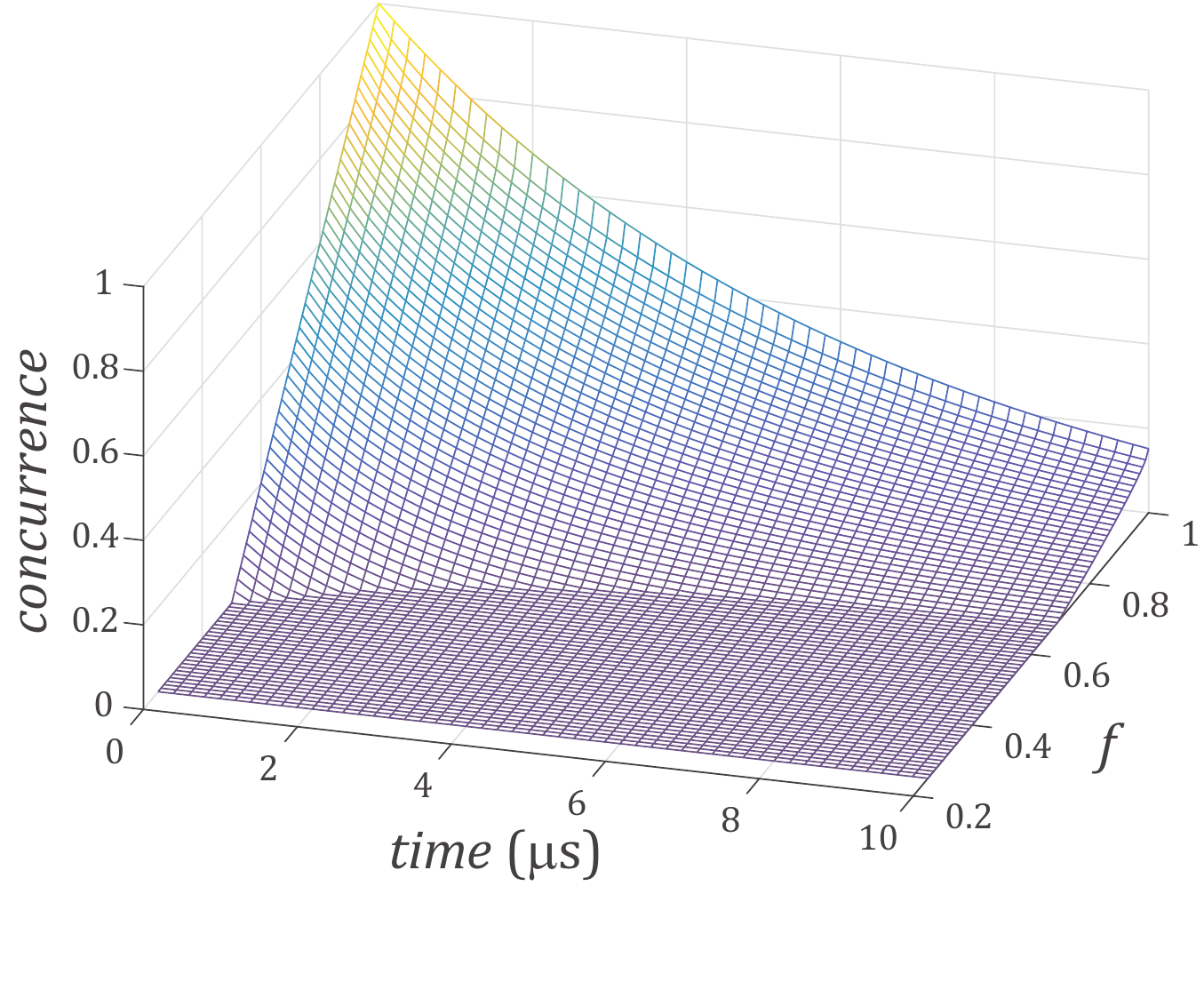}
    \caption{Time evolution of the concurrence for initial Werner state $\rho_W$ at $\lambda / x_2 = 2$.}
    \label{fig:4}
    \end{figure}

    Figure~\ref{fig:4} shows time evolution of the concurrence for that the qubits are initially in the Werner state, and frequency-tuned to $\lambda = 2 x_2$. Without qubit-qubit coupling and common reservoir (collective relaxation), concurrence shows sudden death for $f\leq 0.714$, which is well known for the Werner state.

\subsection{No exchange coupling, finite collective relaxation}
\label{sec:3b}

    When qubit-$a$'s and qubit-$b$'s frequencies are tuned to $\lambda = 1.5 x_2$, the exchange coupling $g_x$ vanishes, and the collective relaxation rate $\Gamma_{col} = -\gamma = -2\pi\times 5$MHz.

    \begin{figure}[bht]
    \includegraphics[width=7cm]{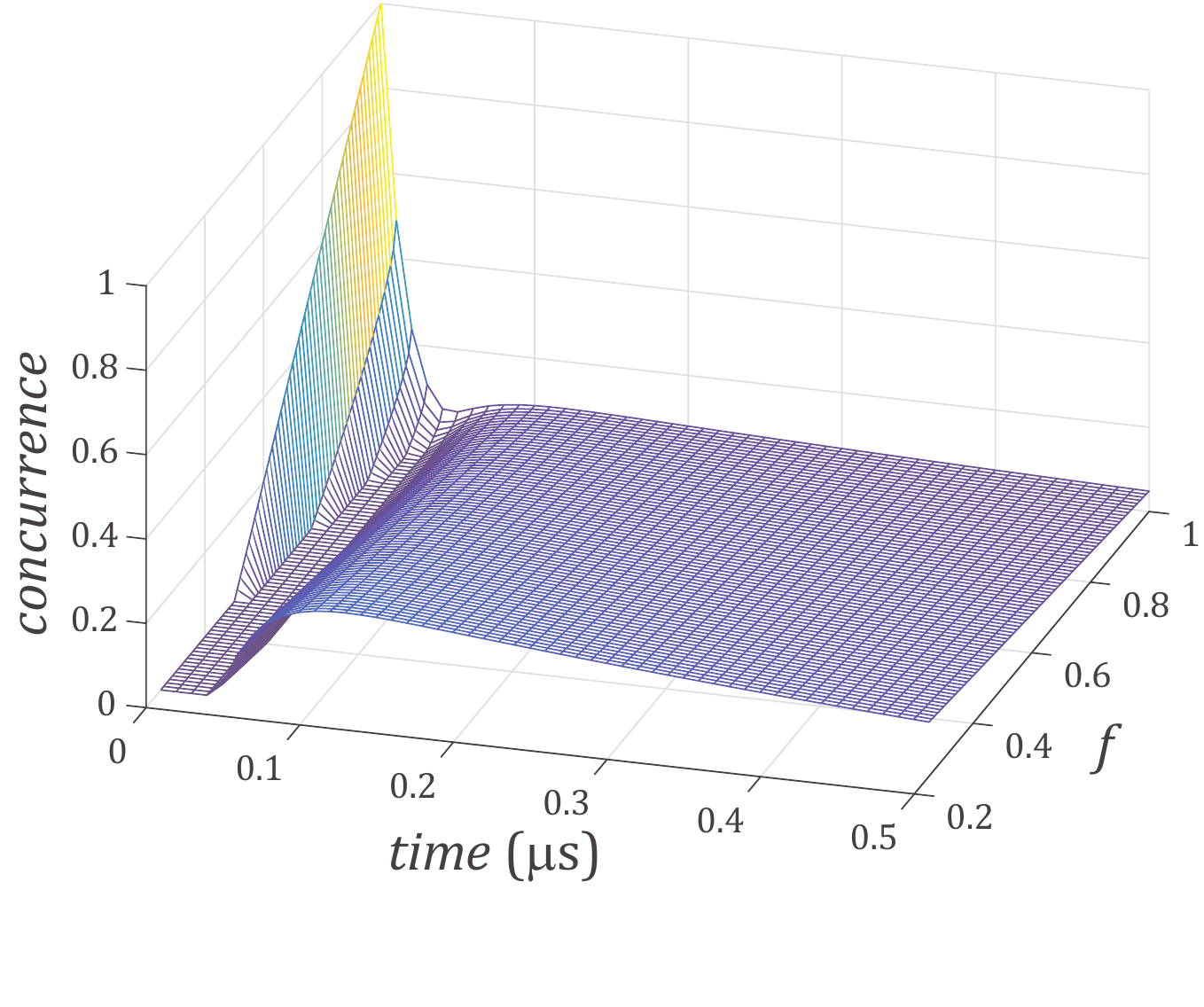}
    \caption{Time evolution of the concurrence for initial Werner state $\rho_W$ at $\lambda / x_2 = 1.5$.}
    \label{fig:5}
    \end{figure}

    The time evolution of the concurrence for the Werner initial state is plotted in Fig.~\ref{fig:5} for this case. The concurrence experiences sudden death even with $f>0.8$, due to the boosted individual relaxations $\Gamma_a$ and $\Gamma_b$. However, because of the existence of the common reservoir, entanglement revives after sudden death. The reviviscent concurrence decays slowly for the rather large individual relaxation rates $\Gamma_a = 2\pi\times 2.53$MHz and $\Gamma_b = 2\pi\times10.03$MHz.

\subsection{Finite exchange coupling, no collective relaxation}

    By tuning the qubits' frequencies to $\lambda = 1.2 x_2$, the exchange coupling strength $g_x \approx -0.866\gamma = -2\pi\times 4.33$MHz, and the collective relaxation rate $\Gamma_{col} = 0$.

    \begin{figure}[bht]
    \includegraphics[width=7cm]{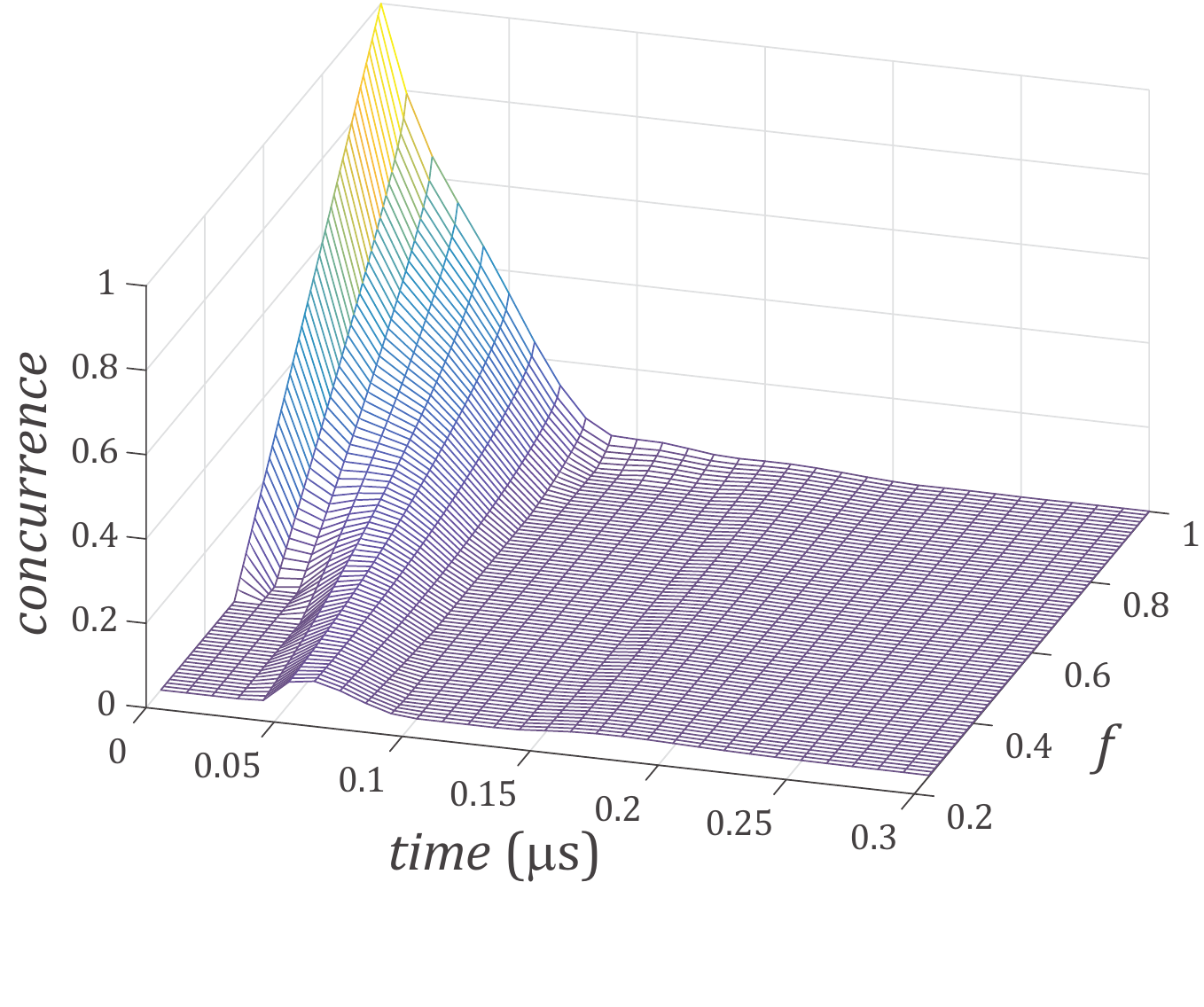}
    \caption{Time evolution of the concurrence for initial Werner state $\rho_W$ at $\lambda / x_2 = 1.2$.}
    \label{fig:6}
    \end{figure}

    In this case, after ESD there is an entanglement revival, as shown in Fig.~\ref{fig:6}. Unlike the case discussed in Sec.~\ref{sec:3b}, here the reviviscent concurrence decays to zero quite fast, even though here the individual relaxation rates $\Gamma_a = 2\pi \times 7.53$MHz and $\Gamma_b = 2\pi\times 0.03$MHz are smaller than those in Sec.~\ref{sec:3b}.

\subsection{Finite exchange coupling, finite collective relaxation}

    By tuning the qubits' frequencies slightly far away from $\lambda = 1.2 x_2$, to {\it e.g.} $\lambda = 1.3 x_2$, a regime in which both the exchange coupling and the collective relaxation are non-zero, $g_x\approx -3.08$MHz and $\Gamma_{col}\approx -4.25$MHz, is reached.

    \begin{figure}[bht]
    \includegraphics[width=7cm]{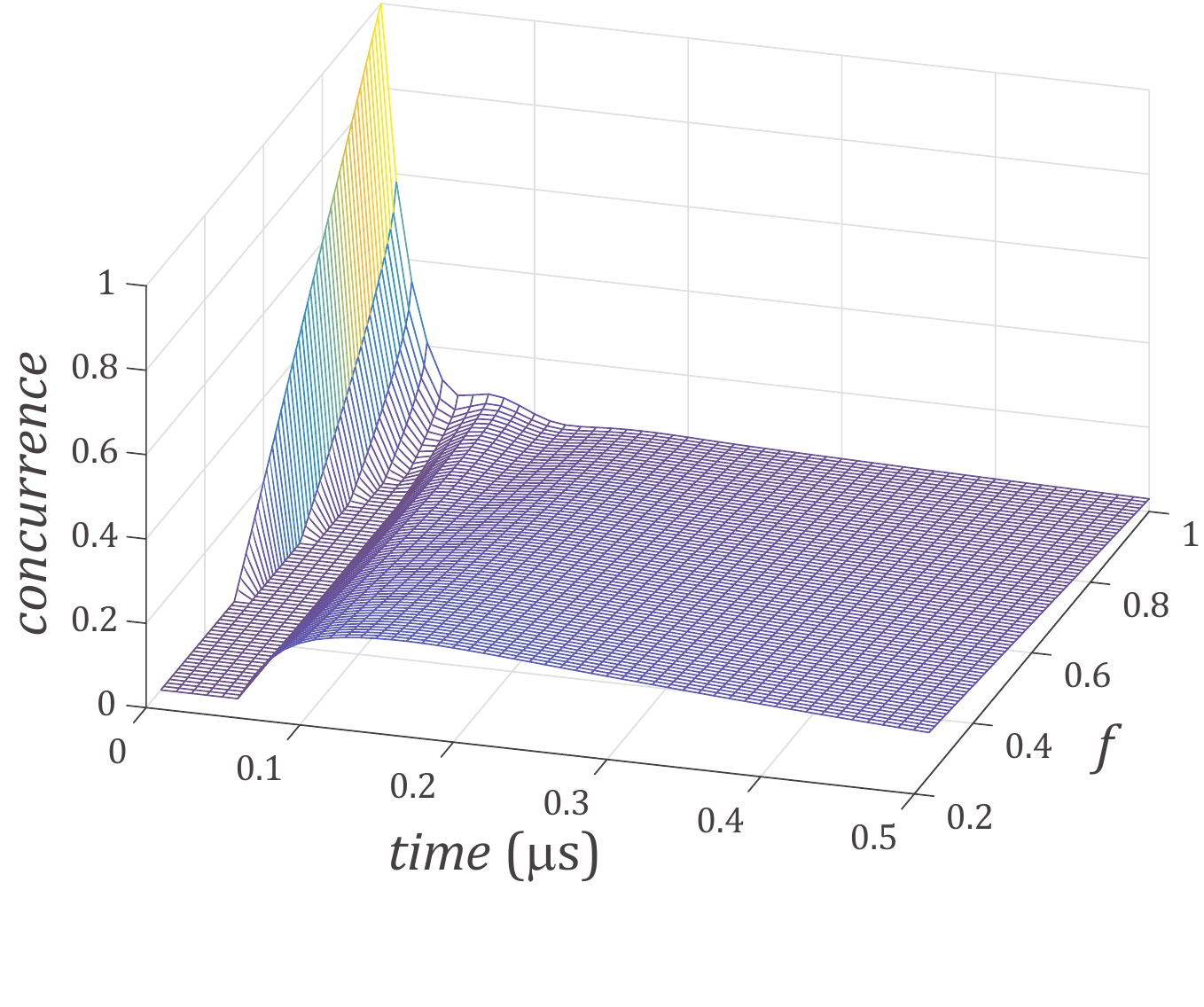}
    \caption{Time evolution of the concurrence for initial Werner state $\rho_W$ at $\lambda / x_2 = 1.3$.}
    \label{fig:7}
    \end{figure}

    The time evolution of the concurrence, as shown in Fig.~\ref{fig:7}, is similar to that shown in Fig.~\ref{fig:5}, even though the individual relaxation rates $\Gamma_a \approx 2\pi\times 5.63$MHz and $\Gamma_b \approx 2\pi\times 3.26$MHz are quite different from those in Sec.~\ref{sec:3b}.

    By comparing the four cases, we may make a summary that for entanglement revival, a common reservoir is more important than a qubit-qubit exchange coupling.

\section{Procedure for observing ESD and revival in waveguide QED}

    Though the Werner state has been frequently used for studying bipartite entanglement, experimentally generating such a state, especially with superconducting qubits, is hard. So far the Werner state has been successfully prepared in optical system \cite{Barbieri2004}. Therefore, here we introduce a new type of 'X' shape state similar to the Werner state
    \begin{equation}
	\rho_{PW}=\frac{1}{4}
	\left(\begin{array}{cccc}
		f/2 & 0           & 0          & 0    \\
		0   & 1+f         & i\sqrt{3}f & 0    \\
		0   & -i\sqrt{3}f & 3f/2       & 0    \\
		0   & 0           & 0          & 3(1-f) 	
	   \end{array}\right), \label{eq:pw_state}
    \end{equation}
    and we call it pseudo-Werner (PW) state.

\subsection{Preparing pseudo-Werner state}

    The PW state can be easily generated in our waveguide QED model system, with an auxiliary qubit (qubit-$c$ in Fig.~\ref{fig:1}). We assume that the coupling between qubit-$a$ and qubit-$b$ (with the strength $g$), as well as the coupling between qubit-$b$ and qubit-$c$ (with the strength $g_{bc}$), have exchange form
    \begin{equation}
    XY = \sigma_-\otimes\sigma_+ + \sigma_+\otimes\sigma_- ,
    \end{equation}
    and that these couplings are switchable. The procedure for generating the PW state is as follows.

    Initially, the three qubits are in a product state
    \begin{equation}
	   \rho_{1} = \rho_{c}\otimes\rho_{b}\otimes\rho_{a},
    \end{equation}
    in which qubit-$a$ is prepared in a mixed state
    \begin{equation}
	\rho_{a}=%
	\left(\begin{array}{cc}
		f & 0    \\
		0 & 1-f  	
	   \end{array}\right) ,
    \end{equation}
    qubit-$b$ and qubit-$c$ are prepared in their excite state and ground state, respectively
    \begin{equation}
	\rho_{b}=%
	\left(\begin{array}{cc}
		0 & 0    \\
		0 & 1  	
	   \end{array}\right), \ \ \ \
    \rho_{c}=%
	\left(\begin{array}{cc}
		1 & 0    \\
		0 & 0  	
	   \end{array}\right) .
    \end{equation}
    A method to prepare qubit-$a$ to the mixed state $\rho_a$ is described in Appendix~\ref{mixed_state} in details.

    Secondly, by switching off the coupling between qubit-$b$ and qubit-$c$, $g_{bc} = 0$, as well as tuning qubit-$a$'s and qubit-$b$'s frequencies to let $\lambda/x_2 =2$, {\it i.e.} $g_{x} = \Gamma_a = \Gamma_b = \Gamma_{col} =0$ (see Fig.~\ref{fig:2} and Fig.~\ref{fig:3}), only the exchange coupling $g$ between qubit-$a$ and qubit-$b$ is non-zero. Using this non-zero exchange coupling to generate the evolution
    \begin{equation}
    \rho_2 = e^{\frac{i\pi}{4}I\otimes XY} \rho_1 e^{\frac{-i\pi}{4}I\otimes XY} ,
    \end{equation}
    we get the state
    \begin{equation}
 	  \rho_{2}=%
 	\left(\begin{array}{cccccccc}
 		0 & 0 & 0 & 0 & 0 & 0 & 0 & 0 \\
 		0 & f/2 & if/2 & 0 & 0 & 0 & 0 & 0 \\
 		0 & -if/2 & f/2 & 0 & 0 & 0 & 0 & 0 \\
 		0 & 0 & 0 & 1 - f & 0 & 0 & 0 & 0 \\
 		0 & 0 & 0 & 0 & 0 & 0 & 0 & 0 \\
 		0 & 0 & 0 & 0 & 0 & 0 & 0 & 0 \\
 		0 & 0 & 0 & 0 & 0 & 0 & 0 & 0 \\
 		0 & 0 & 0 & 0 & 0 & 0 & 0 & 0
 	  \end{array}\right).
    \end{equation}
    Here $I$ denotes the single-qubit identity operator.

    Then, we switch off the coupling $g$ between qubit-$a$ and qubit-$b$, and turn on the coupling $g_{bc}$ between qubit-$b$ and qubit-$c$. The non-zero exchange coupling can generate evolution
    \begin{equation}
    \rho_3 = e^{\frac{i\pi}{6}XY\otimes I} \rho_2 e^{\frac{-i\pi}{6}XY\otimes I} ,
    \end{equation}
    \begin{widetext}
    and produce a three-qubit state
 	\begin{equation}
 		\rho_{3}=%
 		\left(\begin{array}{cccccccc}
 			0 & 0 & 0 & 0 & 0 & 0 & 0 & 0 \\
 			0 & f/2 & i(1/4)\sqrt{3}f & 0 & f/4 & 0 & 0 & 0 \\
 			0 &-i(1/4)\sqrt{3}f & 3f/8 & 0 & -i(1/8)\sqrt{3}f & 0 & 0 & 0 \\
 			0 & 0 & 0 & -(3/4)(-1 + f) & 0 & i(1/4)\sqrt{3}(-1 + f) & 0 & 0 \\
 			0 & f/4 & i(1/8)\sqrt{3}f & 0 & f/8 & 0 & 0 & 0 \\
 			0 & 0 & 0 & -i(1/4)\sqrt{3}(-1 + f) & 0 & (1 - f)/4 & 0 & 0 \\
 			0 & 0 & 0 & 0 & 0 & 0 & 0 & 0 \\
 			0 & 0 & 0 & 0 & 0 & 0 & 0 & 0 		
 		 \end{array}\right).
 	  \end{equation}
    \end{widetext}

    By partially tracing out the auxiliary qubit-$c$ from $\rho_3$, the subsystem of qubit-$a$ and qubit-$b$ is finally in the PW state Eq.~(\ref{eq:pw_state}).

\subsection{Time evolution of entanglement for initial pseudo-Werner state}

    Before studying the dynamics of the entanglement for pseudo-Werner state, let us consider the concurrence of $\rho_{PW}$ itself. For $\rho_{PW}$, the concurrence is simply (see Appendix~\ref{x_state})
    \begin{equation}
    C = 2\max[0,F,G] ,
    \end{equation}
    where
    \begin{eqnarray}
    & F = \sqrt{3} [2f - \sqrt{2f(1-f)}]/8, \nonumber \\
    & G = -\sqrt{3f(1+f)/32} .
    \end{eqnarray}
    Since $0.25\leq f \leq 1$, obviously $G<0$ for all values of $f$. Thus the concurrence is determined by $F$. The solution of the equation $F = 0$ is $f = 1/3$, which means that for $f\leq 1/3$ the pseudo-Werner state is separable. The maximum concurrence $C = \sqrt{3}/2\approx 0.866$ is reached at $f = 1$.

\begin{widetext}
    \begin{figure}[bht]
    \includegraphics[width=14.9cm]{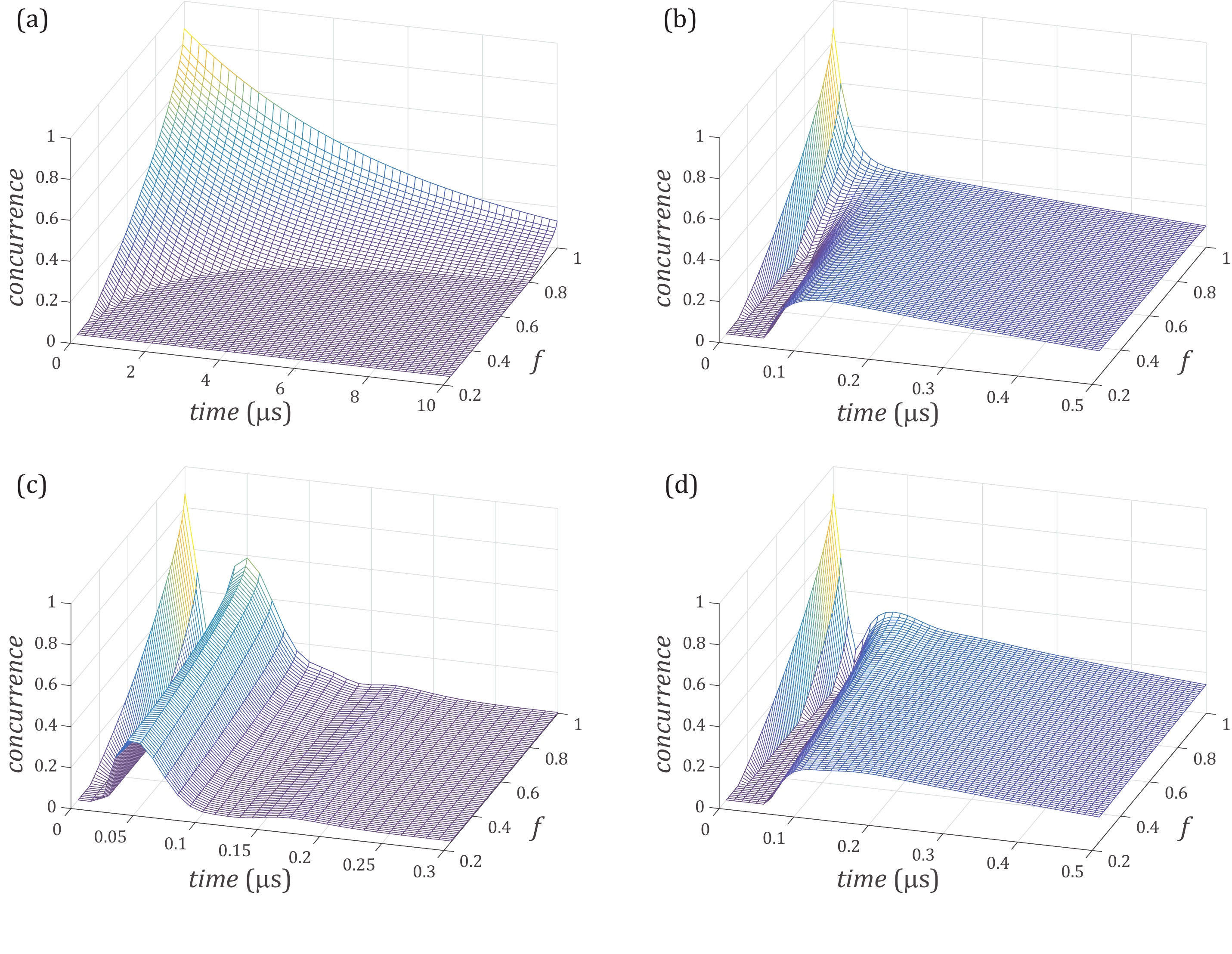}
    \caption{Time evolution of the concurrence for initial pseudo-Werner state $\rho_{PW}$ at (a) $\lambda / x_2 = 2$, (b) $\lambda / x_2 = 1.5$, (c) $\lambda / x_2 = 1.2$, and (d) $\lambda / x_2 = 1.3$.}
    \label{fig:8}
    \end{figure}
    \end{widetext}

    We then calculate time evolutions of the concurrences for the pseudo-Werner initial state with the same parameters and wavelengths as those in Sec.~\ref{sec:3}. The concurrences are plotted in Fig.~\ref{fig:8}. Comparing the concurrences in Fig.~\ref{fig:8} with those in Figs.~{\ref{fig:4}}-\ref{fig:7}, we can find that the entanglement dynamics of the pseudo-Werner state behaves very similar to that of the Werner state. Due to the simplicity of preparing the pseudo-Werner state, $\rho_{PW}$ can become an ideal experimental toy for investigating bipartite entanglement.

\subsection{Design of waveguide QED}

    Superconducting qubits' transition frequencies are in GHz range. Due to the limitation of bandwidth of the high performance cryogenic {\it rf} components such as low noise cryogenic amplifiers and circulators, superconducting qubit readout resonator for dispersive measurements is normally designed between $4$GHz and $8$GHz. In this waveguide QED system, let us assume that the readout resonators for qubit-$a$ and qubit-$b$ are at about $8$GHz. Then in order to be able to perform dispersive readout of the qubits, the maximum transition frequency should be no higher than $7$GHz for both qubits.

    The qubit frequency dependent wavelength
    \begin{equation}
    \lambda = 2\pi v_{ph} / \omega_a ,
    \end{equation}
    where $v_{ph}$ is the phase velocity in the waveguide. We use a conventional CPW design for the waveguide. Considering that the substrate is sapphire ($\varepsilon_r =9.8$), and that the metal parts of the CPW are fabricated by aluminum (since the kinetic inductance of aluminum is small), the characteristic impedance $Z_0$ and the phase velocity $v_{ph}$ are calculated by following Ref.~\cite{Simons2001}.

\begin{figure}[bht]
    \includegraphics[width=7cm]{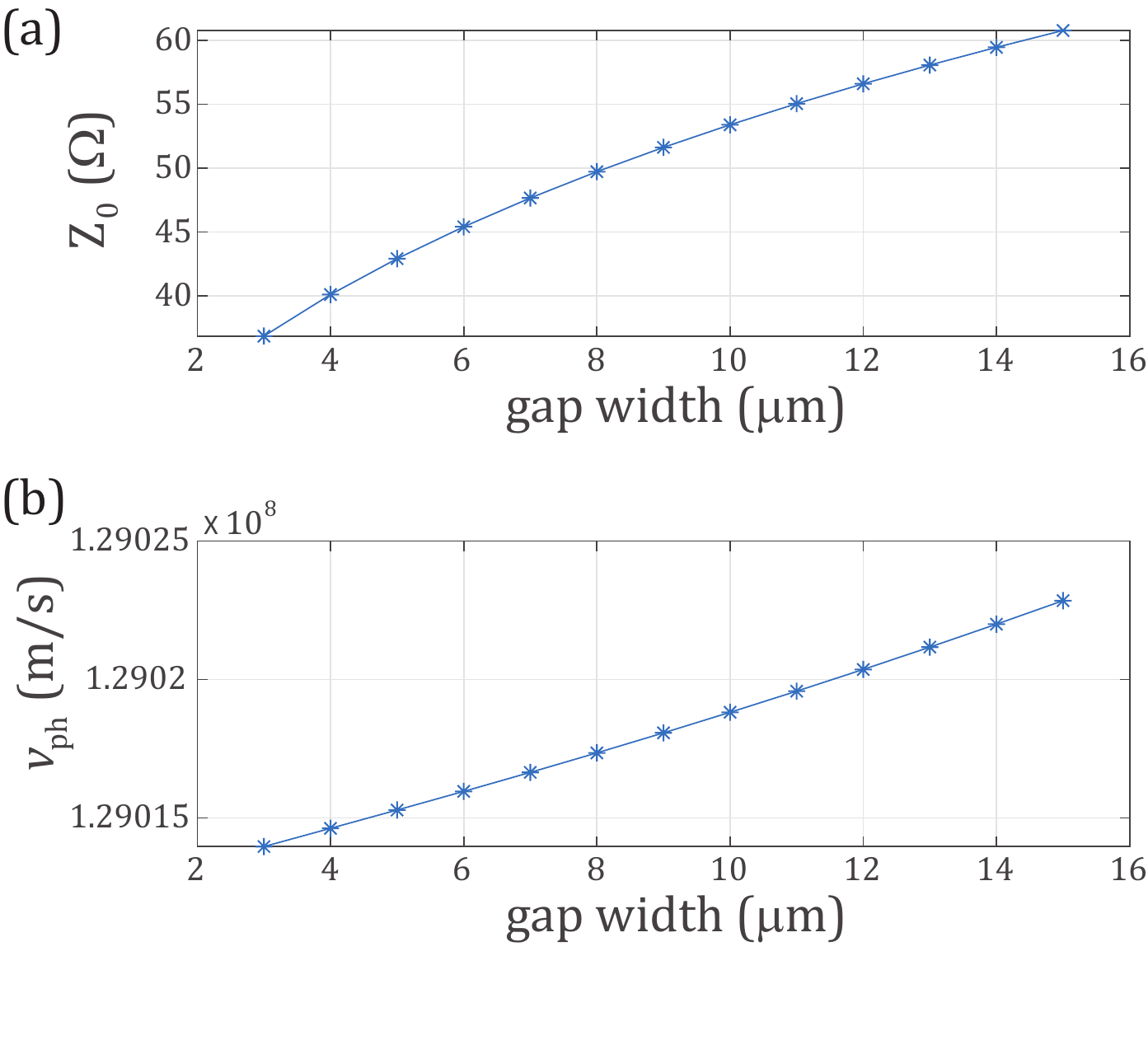}
    \caption{(a) Characteristic impedance $Z_0$ versus vacuum gap width. (b) Phase velocity $v_{ph}$ versus vacuum gap width.
    }
    \label{fig:vph}
    \end{figure}

    By fixing the centre conductor width to $20\mu$m, as shown in Fig.~\ref{fig:vph}, when the vacuum gap width is $8\mu$m the characteristic impedance of the waveguide is matched to $50\Omega$, and the phase velocity $v_{ph}\approx 1.29017\times 10^8$m$/$s. As mentioned above the maximum qubit frequency $\omega_a = 2\pi\times 7$GHz, corresponding to a wavelength $\lambda \approx 18.4$mm. Thus we can design the distances $x_1 = 9.2$mm and $x_2 = 18.4$mm, so that $\lambda/x_2 = 1$ (see Fig.~\ref{fig:2} and Fig.~\ref{fig:3}) can be reached. For $\lambda / x_2 = 3$, the qubit frequency needs to be tuned to $\omega_a \approx 2\pi\times 2.3$GHz, which might be too low for a dual-junction transmon qubit, however, $\lambda / x_2 = 2.4$ should be feasible at qubit frequency $\omega_a \approx 2\pi\times 3$GHz.

\section{Conclusion}

    In conclusion, we have shown the phenomena of ESD and entanglement revival for different 'X' shape initial states under individual and collective decays utilizing a superconducting waveguide QED system. We find for both the Werner state and the pseudo-Werner state, the evolution of concurrence is dependent heavily on the wavelength determined by qubits' transition frequencies. We later demonstrate a detailed experimental procedure to observe ESD and revival in the superconducting waveguide QED system. The results may contribute to further understanding of the evolution of entanglement in dissipative quantum systems.	

\begin{acknowledgments}
    J.L. acknowledges support from the Key-Area Research and Development Program of Guang-Dong Province (Grant No. 2018B030326001), the National Natural Science Foundation of China (Grant No. 11874065), the Guangdong Provincial Key Laboratory (Grant No.2019B121203002), the Science,Technology and Innovation Commission of Shenzhen Municipality (KYTDPT20181011104202253), the ShenzhenHong Kong Cooperation Zone for Technology and Innovation (HZQB-KCZYB-2020050).
\end{acknowledgments}
				
\appendix

\section{Concurrence calculation for 'X' shape states}
\label{x_state}

    At any time $t$, the bipartite density matrix has the form

    \begin{eqnarray}
	\rho(t) =\left(\begin{array}{cccc}
		a(t)        & 0      & 0    & w (t)\\
		0           & b(t)   & z(t) & 0         \\
		0           & z(t)^* & c(t) & 0         \\
		w(t)^* & 0      & 0    & d(t)        	
	   \end{array}\right).
    \label{density matrix}
    \end{eqnarray}
    Substituting Eq.~(\ref{density matrix}) into Eq.~(\ref{master equation}), we can obtain the following kinetic equations of the non-zero density matrix elements

    \begin{eqnarray}
	\dot w(t) = && -\frac{1}{2}w(t) [2(\gamma+\gamma_{nr})+2i(\omega_a+ \omega_b) \nonumber \\
&& \ \ \ \ \ \ \ \ \ \ \ + \gamma(\cos\phi+\cos3\phi) + i\gamma(\sin\phi+\sin3\phi)],\nonumber\\
	\dot z(t) = && \frac{1}{2}\{-2z(t)(\gamma+\gamma_{nr})-2iz(t)(\omega_a-\omega_b) \nonumber\\
	&&\ \ \ +[2d(t)-b(t)-z(t)-c(t)]\gamma\cos\phi\nonumber\\
	&&\ \ \ +[2d(t)-b(t)-c(t)]\gamma\cos2\phi-z(t)\gamma\cos3\phi\nonumber \\
	&&\ \ \ +i[b(t)-c(t)-z(t)]\gamma\sin\phi\nonumber\\
	&&\ \ \ +i[b(t)-c(t)]\gamma\sin2\phi+iz(t)\gamma\sin3\phi\},\nonumber\\
\dot a(t) = &&
[b(t)+c(t)](\gamma+\gamma_{nr})+[c(t)+z(t)+z(t)^*]\gamma\cos\phi\nonumber\\
	&& +[z(t)+z(t)^*]\gamma\cos2\phi+b(t)\gamma\cos3\phi,\nonumber\\
\dot b(t) = &&
(\gamma+\gamma_{nr})[d(t)-b(t)] \nonumber \\
&& + \{d(t)-\frac{1}{2}[z(t)+z(t)^*]\}\gamma\cos\phi\nonumber\\
&& - \frac{1}{2}[z(t)+z(t)^*]\gamma\cos2\phi-b(t)\gamma\cos3\phi\nonumber\\
&& + \frac{1}{2}i\gamma[z(t)-z(t)^*](\sin\phi+\sin2\phi),\nonumber\\
	\dot c(t) = &&
(\gamma+\gamma_{nr})[d(t)-c(t)] \nonumber \\
&& - \{c(t)+\frac{1}{2}[z(t)+z(t)^*]\}\gamma\cos\phi\nonumber\\
&& - \frac{1}{2}[z(t)+z(t)^*]\gamma\cos2\phi+d(t)\gamma\cos3\phi\nonumber\\
&& - \frac{1}{2}i\gamma[z(t)-z(t)^*](\sin\phi+\sin2\phi),\nonumber\\
	\dot d(t)=&&-d(t)[2(\gamma+\gamma_{nr})+\gamma(\cos\phi+\cos3\phi)]. \nonumber
    \end{eqnarray}

    The concurrence is given by

    \begin{equation}
	   C(\rho) = 2\max[0,F,G] ,
    \end{equation}
    where
    \begin{eqnarray}
    F = |z| - \sqrt{ad}, \ \ \ \ G = |w| - \sqrt{bc} . \nonumber
    \end{eqnarray}

\section{Generating mixed state}
\label{mixed_state}

    Since normally a superconducting qubit has an intrinsic 'non-radiative' decay with rate $\gamma_{nr}$, this intrinsic decay can be utilized to generate the mixed state
    \begin{equation}
	\rho_{a}=%
	\left(\begin{array}{cc}
		f & 0    \\
		0 & 1-f  	
	   \end{array}\right)
    \label{eq:mix1}
    \end{equation}
    of qubit-$a$. Assuming initially the qubit is on its ground state, an on-resonance (with the qubit's transition frequency) Rabi pulse much longer than $1/\gamma_{nr}$ drives the qubit into a mixed state
    \begin{equation}
	\rho_{m} \approx
	\left(\begin{array}{cc}
		1/2 & 0    \\
		0 & 1/2  	
	   \end{array}\right) .
    \label{eq:mix2}
    \end{equation}

\begin{figure}[bht]
        \includegraphics[width=7cm]{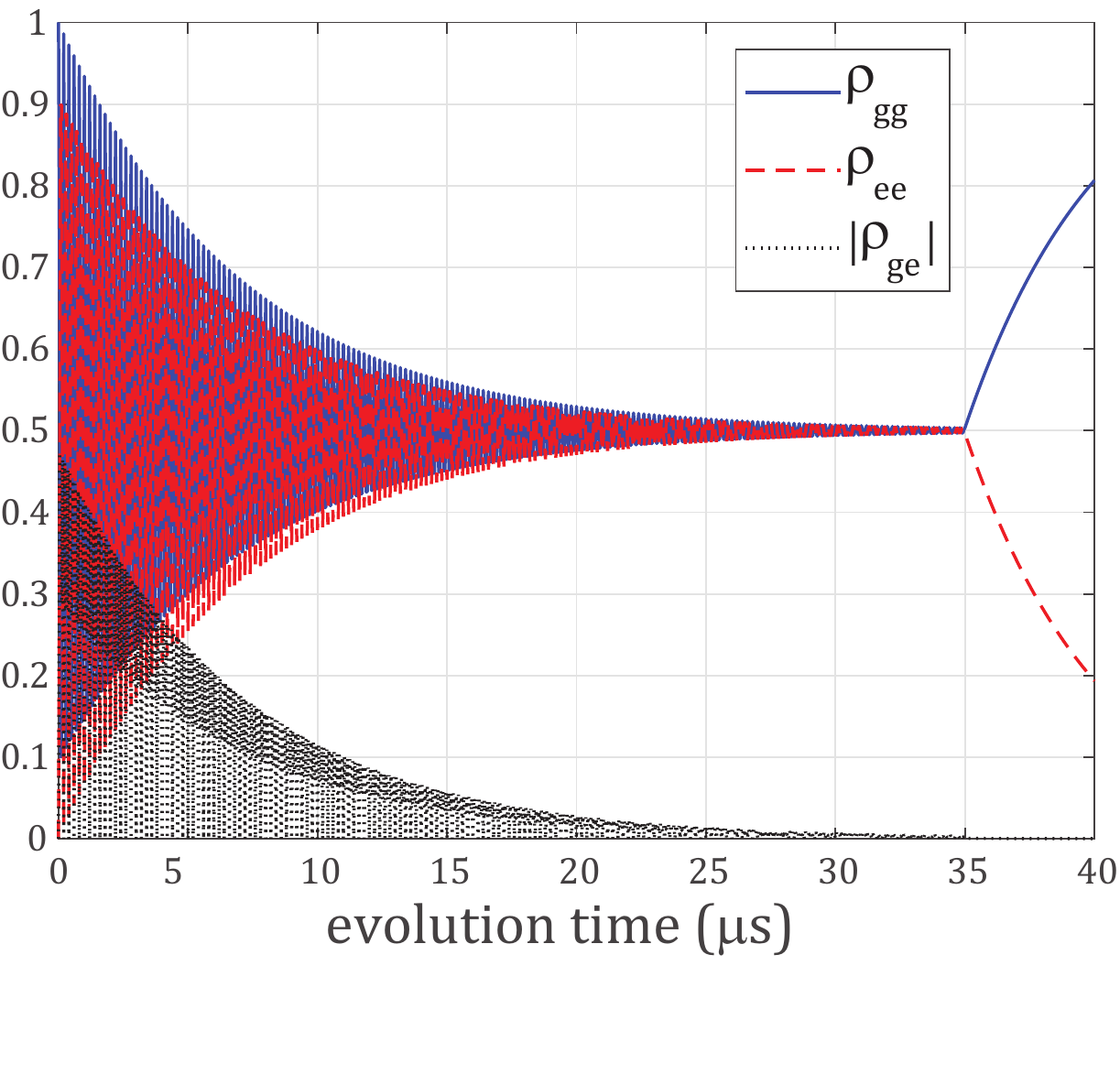}
    \caption{Time evolution of the density matrix elements. The on-resonance Rabi pulse has a simple square envelope, which is switched on at $t = 0$ and switched off at $t = 35\mu$s.
    }
    \label{fig:mix}
    \end{figure}

    We numerically calculate the time evolution of the density matrix diagonal elements $\rho_{gg}$ (the ground state population), $\rho_{ee}$ (the excited state population), and the absolute value of the off-diagonal element $\rho_{eg}$, by taking $\gamma_{nr} = 2\pi\times 0.03$MHz and Rabi frequency $\Omega = 2\pi\times 30$MHz. As shown in Fig.~\ref{fig:mix}, a $35\mu$s long Rabi pulse can drive the qubit to the mixed state Eq.~(\ref{eq:mix2}).

    After switching off the Rabi pulse, the off-diagonal density matrix elements remain zero, while the diagonal element $\rho_{ee}\  (\rho_{gg})$ decays (increases) exponentially. Thus, depending on how long time to wait after the Rabi pulse, arbitrary fidelity factor $f\geq 0.5$ can be obtained. For example, $f = 0.8$ is reached $5\mu$s after the Rabi pulse, as shown in Fig.~\ref{fig:mix}. Fidelity factor $f<0.5$ can be realized by simply applying an extra $\pi$-pulse to flip the population.

\bibliography{Entanglement_properties_of_superconducting_qubits_coupled_to_a_semi-infinite_transmission_line}

\end{document}